\documentclass[twocolumn,pre,showpacs]{revtex4}
\usepackage{epsfig}
\usepackage{graphicx}
\usepackage{epsfig}
\usepackage{dcolumn}
\usepackage{bm}
\begin{document}

\title{Glass Phenomenology from the Connection to Spin Glasses }
\author{M. Tarzia and M. A. Moore}
\affiliation{School of Physics and Astronomy, University of Manchester, 
Manchester M13 9PL, United Kingdom}
\date{\today}

\begin{abstract}

Typical features of glass phenomenology such as the Vogel-Fulcher law,
the  Kauzmann paradox  and the  Adam-Gibbs relationship  are shown to follow
from the recently  discovered mapping of glasses to  Ising 
spin glasses in a  magnetic field. There seems to  be sufficient universality
near the  glass transition  temperature $T_g$ such  that study  of the
spin glass  system   can  provide   semi-quantitative   results  for
supercooled liquids.

\end{abstract}
\pacs{64.70.Pf, 75.10.Nr, 75.50.Lk}
\maketitle

\section{introduction}

Under fast enough cooling or densification, materials which are as diverse as
 molecular
and polymeric liquids, colloidal suspensions, granular assemblies and molten
mixtures of metallic atoms, may form glasses~\cite{DeSt01}. These are
 amorphous states that may be
characterized mechanically as a solid, but
lack long-range crystalline order.
Despite all the work devoted to the subject,
the mechanisms responsible for the vitrification processes
are not well understood, and the transition to the glassy state
remains one of the most controversial problems  in
condensed matter physics.

In a recent paper~\cite{MoYe06}, using an effective potential method,
a replica formalism has been set up to describe supercooled liquids.
This approach shows that the thermodynamics  of these systems near 
their glass transition temperature $T_g$ is equivalent, in  the sense of 
 ``universality  classes'', to
that of  Ising spin glasses in  a magnetic field  $h$~\cite{MoYe06}. 
 Spin-glass droplet scaling ideas~\cite{mike, FiHu} were used to
discuss the consequent expected glass phenomenology.
 This 
approach
would be appropriate if the lengthscales of cooperatively 
rearranging regions near 
$ T_g$  were many times larger than 
 the intermolecular separation. In fact, recent studies~\cite{length} 
have indicated that this lengthscale is rather modest and only a few
 intermolecular distances. As a consequence, glasses are
not really in the regime where droplet 
scaling ideas are appropriate.  We shall therefore
 examine in this 
paper the glass phenomenology which arises when the 
 correlation length is not  large but instead is in 
 the  precursor regime to  the droplet
scaling limit.
Rather to our surprise, we are able to find in this regime 
all the characteristic features 
of  glass  phenomenology  such  as  the  Vogel-Fulcher 
 relation~\cite{DeSt01},  the
Kauzmann  temperature~\cite{Ka48}, the  Adam-Gibbs relation~\cite{AdGi}, etc...

We shall study in particular  the Edwards-Anderson (EA) Ising  spin glass 
model~\cite{EAmodel} 
in the presence of an external magnetic field $h$ both in one dimension ($1d$)
and in three dimensions ($3d$). 
The behavior in both dimensions is similar, but the
$1d$ case can be studied more thoroughly as
its equilibrium properties  can be determined exactly by means of a
  renormalization
group 
approach, and its dynamical properties are accessible via 
  Monte Carlo simulations.
Even in the $1d$ case, the model is able to
mimic most of 
the experimental observations on supercooled liquids.
 An apparent 
Kauzmann paradox~\cite{Ka48} 
is found, accompanied by a growing (but still modest at the fields
which we use) lengthscale $\xi$ and by an
apparent divergence of the relaxation time as in the Vogel-Fulcher (VF)
equation  with $T_{VF} = T_K$~\cite{DeSt01}. There is thus
an apparent thermodynamical and dynamical singularity at a finite
temperature $T_K$, but it  is not a true transition. 
$T_K$ 
is just a crossover temperature such that when  $T \lesssim T_K$ 
the growth of $\xi$ as $T$ decreases has largely ceased.
   The $3d$ 
case cannot be solved exactly but has been studied within the
Migdal-Kadanoff approximation (MKA)~\cite{kadanoff,MKA},
 and a similar glass phenomenology emerges. But there are some significant
difference between the $1d$ and $3d$ cases, due to the fact in $1d$ 
there is no finite temperature spin glass transition, but only a diverging
lengthscale as $T\rightarrow 0$ when $h=0$, whereas in $3d$, there is in zero field
a finite temperature transition and so in order to have a lengthscale of only a few
intermolecular diameters at low temperatures, a large field has to be applied.
 Our work within the MKA in $3d$
 does not provide a quantitatively accurate picture,
 but the results  are so encouraging 
  that it would seem worthwhile to attempt to get more quantitative results,
 probably by use of
Monte Carlo methods. Unfortunately glass timescales are so long compared to 
molecular collision timescales that realistic simulations
 will be challenging.

The investigations presented here suggest a new framework to
understand glass behavior: features of the intermolecular potential and the 
density 
 determine the value of the field $h$ and the temperature scale, but once
 these are fixed  there is sufficient universality left  near 
 the glass transition $ T_g$ that the mapping
 to Ising spin glasses in a field  provides a semi-quantitative account
 of both the 
thermodynamic properties of glasses  and those dynamical features which 
can be undestood in terms of flipping and cooperative rearranging of spin
domains of linear extent $\xi$ sitting in a random effective magnetic field, 
requiring  free energy activation over  barriers.

This paper is organized as follows: in Sec.~\ref{sec:1d} we solve the
EA Ising spin glass in an external magnetic field in one dimension
by using a decimation
approach and we study its dynamical properties by performing Monte Carlo
simulations; we then discuss the connection to the phenomenology of glasses.
In Sec.~\ref{sec:3d} we examine the three dimensional case, by means of
the Migdal-Kadanoff approximation. Finally, in Sec.~\ref{sec:concl}, we 
present some concluding remarks.

\section{The EA Ising Spin Glass in one dimension} \label{sec:1d}
The EA Ising
spin glass Hamiltonian in $1d$ 
in the presence of a uniform external magnetic field reads:
\begin{equation} \label{eq:hea}
{\cal H} = - \sum_i J_i \sigma_i \sigma_{i+1} - h \sum_i \sigma_i,
\end{equation}
where the  spins $\sigma_i$ can take values $\pm 1$, and the 
nearest-neighbor couplings $J_i$ are independent of each other and Gaussianly
distributed with zero mean and standard deviation $J$. In principle
 $h$ could be a function of temperature but we shall regard it as 
a temperature independent constant, whose magnitude is chosen both in $1d$ and
 $3d$  so that the
 low-temperature spin glass correlation
 length $\xi$ is of the order of a few lattice spacings and so is comparable to
the glass correlation length of real glasses at $T_g$~\cite{length}.
We evaluate the free energy of the system by using an iterative real-space
renormalization group technique~\cite{kadanoff,MKA}. 
It consists in tracing out every other spin in the system, thereby  
generating  new effective interactions between the remaining spins
which sit in new  magnetic fields:
\begin{eqnarray} \label{eq:1drg}
\!\!\!\!\!\!\!\!\!\!\!\!&& \sum_{\sigma_{i+1} = \pm 1} e^{ \beta \left [
J_{i}^{(n)} \sigma_{i} \sigma_{i+1}
+ J_{i+1}^{(n)} \sigma_{i+1} \sigma_{i+2} + \sum_{j=i}^{i+2} h_{j}^{(n)}
\sigma_{j} \right]} \\
\nonumber
&& \,\,\, = e^{ W_{i+1}^{(n+1)} + \beta \left[
J_i^{(n+1)} \sigma_i \sigma_{i+2} + h_i^{(n+1)} \sigma_i
+ h_{i+2}^{(n+1)} \sigma_{i+2} \right]},
\end{eqnarray}
At the $n$-th step in the decimation process,
$J_i^{(n)}$ and $h_i^{(n)}$, have
 probability distributions, which
evolve with the iteration. In $1d$, the EA model has a genuine
critical point at $(T=0, h=0)$, corresponding to a non-trivial
fixed point of the recursion relations. Conversely, at any finite temperature
and magnetic field, the system evolves toward a trivial ``random paramagnetic''
fixed point: the variance of the effective couplings decreases under
iteration and approaches zero, 
whereas the effective magnetic fields have a
distribution which 
approaches a Gaussian, with mean $h$ (i.e., the initial value of
the magnetic field), and variance $\sigma_h (T,h)$.

\begin{figure}[ht]
\vspace{-0.1cm}
\begin{center}
\psfig{figure=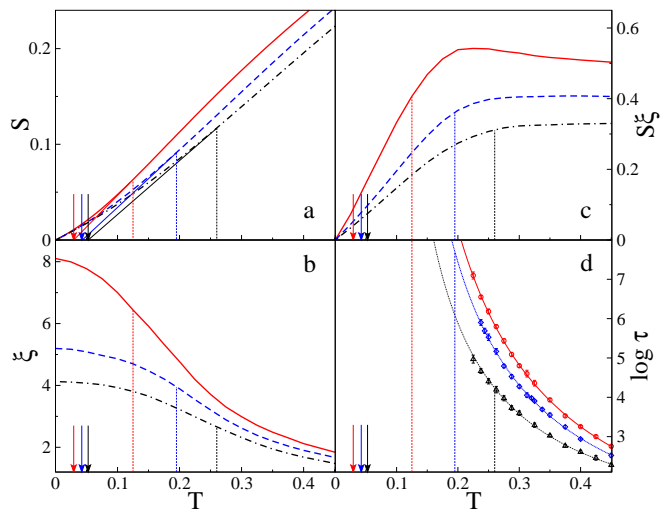,scale=0.31,angle=270}
\end{center}
\vspace{-0.4cm}
\caption{One dimensional EA model in a field 
for $h=0.05$ (continuous line and circles), $h=0.125$ (dashed line
and diamonds) and $h=0.2$ (dashed-dotted line and triangles).
{\bf a)} Entropy per spin, $S$;
{\bf b)} Equilibrium correlation length, $\xi$;
{\bf c)} Adam-Gibbs relation: temperature dependence of $S \xi$;
{\bf d)} Logarithm of the relaxation time, $\tau$, as a function of the 
temperature. The curves correspond to the  
Vogel-Fulcher fits, Eq.~(\ref{eq:vtf}), with $T_{VF} = T_K$.
The arrows indicate the Kauzmann temperatures, $T_K$, whereas the vertical
dashed lines correspond to the crossover temperatures, $T^{\star}$.}
\label{fig:1d}
\vspace{-0.3cm}
\end{figure}

For each realization of the quenched disorder, 
the free energy, $f_J$, 
can be determined exactly by summing the spin-independent terms, $W_i^{(n)}$,
which are generated at each step of the decimation~\cite{MKA}.
Once the average over the disorder
is taken, $f = [f_J]_J$, 
the entropy density is obtained using
$S = -\partial f/\partial T |_h$.
In Fig.~\ref{fig:1d}a  the entropy is 
plotted versus the temperature for three different value of the magnetic field,
$h = 0.05$, $0.125$ and $0.2$.
The figure shows a temperature range in which the entropy decreases
linearly, and would be extrapolated to vanish at   
$T_K (h)$ as
\begin{equation} \label{eq:1dentr}
S \simeq k_B c(h) [T - T_K (h)],
\end{equation} 
demonstrating that
the model has a ``Kauzmann paradox'' similar to that observed in
supercooled liquids.
However, below a crossover temperature, $T^{\star} (\lesssim T_g)$, 
the entropy deviates from linearity  
and does not vanish completely except at $T=0$.

By computing the derivative of the free energy with respect to the
variation of the magnetic field on two different sites, it is possible to \
evaluated the equilibrium connected correlation function:
\begin{equation}\label{eq:corr}
\langle \sigma_i \sigma_{i+l} \rangle_c^2 = \left ( T^2 \, 
\frac{\partial^2 \ln Z}
{\partial (\delta h_i) \partial (\delta h_{i+l})} \bigg|_{\delta h_i = \delta 
h_{i+l} = 0} \right)^2.
\end{equation}
>From the exponential decay of the correlation function,
$[ \langle \sigma_i \sigma_{i+l} \rangle_c^2 ]_J \sim \exp (- l / \xi )$, 
we extract the equilibrium correlation length of the system, $\xi$,
plotted in Fig.~\ref{fig:1d}b for the same values of the
magnetic field as before.  
$\xi$ increases as the temperature
is decreased but at low enough temperatures ($T \lesssim T^{\star}$), 
it bends over and approaches
a finite value at $T=0$, proportional
to $h^{-2/3}$, as predicted by the droplet picture~\cite{mike,FiHu}
 on equating the energy to flip a domain of size $\xi$ 
 to the magnetic field
energy which could be gained, $J\xi^{\theta}\sim h\xi^{d/2} (\sim k_BT_K)$,
 and for $d=1$, $\theta=-1$.

\begin{figure}[ht]
\vspace{-0.1cm}
\begin{center}
\psfig{figure=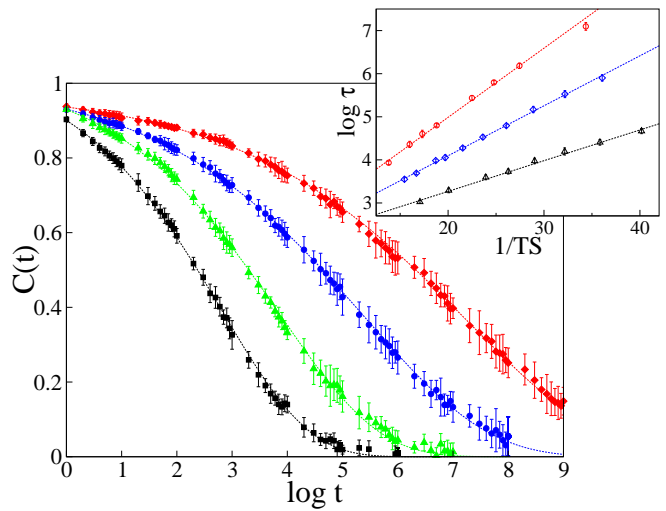,scale=0.31,angle=270}
\end{center}
\vspace{-0.4cm}
\caption{{\bf Main frame:} 
Time dependent
spin-spin auto-correlation function, $C(t)$, as a
a function of $\log t$ for $h = 0.05$ and $T = 0.45$ (squares),
$T = 0.35$ (triangles), $T = 0.275$ (circles) and $T = 0.225$
(diamonds). The data are averaged over $16$-$32$ independent
realisations of the disorder and over $t_w$.
The dashed line correspond to stretched exponential
fits $C(t) \sim \exp[-(t/\tau)^{\beta}]$ with $0.2 < \beta < 0.3$.
{\bf Inset:} Adam-Gibbs relation:
Logarithm of the relaxation time, $\tau$, as a function of $1/TS$ for 
$h=0.05$ (circles), $h=0.125$ (diamonds) and $h=0.2$ (triangles).
The straight lines are guides for the eye.}
\label{fig:relax}
\vspace{-0.3cm}
\end{figure}

In order to establish a connection
with the dynamical features of glass-forming liquids,
we have studied the dynamical properties of the $1d$ model
by performing Monte Carlo 
simulations of a system of $1024$ spins.  We have computed the spin-spin
 cumulant
auto-correlation function, defined as:
\begin{equation}
C(t, t_w) = \left [ \frac{1}{N}  
\sum_i \langle \sigma_i (t + t_w) \sigma_i (t) \rangle_c
\right]_J.
\end{equation}
For large enough waiting times, $t_w$, the system reach stationarity, 
characterised by time transitional invariance, i.e., $C(t,t_w) = C(t)$.
Although the mapping 
established in Ref.~\cite{MoYe06} is explicit only for equilibrium
quantities and might not extend to dynamical features, we find that, 
in analogy with glass-formers, 
$C(t)$ is very well fitted by a stretched exponential form,
$C(t) \sim \exp [- (t/\tau)^{\beta}]$, where $\tau (T,h)$ is the 
system relaxation time. The spin-spin auto-correlation function is plotted
in Fig.~\ref{fig:relax} for several values of the temperature.

The relaxation time is plotted in Fig.~\ref{fig:1d}d for 
$h = 0.05$, $0.125$ and $0.2$ as a function of the temperature. Similarly to
what happens in supercooled liquids~\cite{DeSt01}
a Vogel-Fulcher law, 
\begin{equation} \label{eq:vtf}
\log \tau = \log \tau_0 + \frac{D \, T_{VF}}{T - T_{VF}},
\end{equation}
is able to fit quite accurately the data for each value of the magnetic field
(over $4$-$5$ decades),
with the VF temperature, $T_{VF}$, set equal to the Kauzmann
 temperature, 
$T_{K}$~\cite{DeSt01}. 
Again, the dynamical singularity  
is only apparent, since 
the relaxation time 
diverges only at $T=0$ according to
an Arrhenius law. At low enough temperature $\tau$ deviates from the VF law. 
Such a departure starts to emerge for $h=0.2$ when $T \lesssim 0.25$. 
Interestingly, the onset of the deviation from the VF fit  
seems to coincide with the
crossover temperature $T^{\star}$ 
at which the entropy deviates from linearity. 
This behavior is 
consistent with the Adam-Gibbs (AG) relation~\cite{AdGi},
$\log \tau = A_{AG} + B_{AG}/T S(T)$,
which holds in the 
temperature range explored by the simulations,
as shown in the inset of Fig.~\ref{fig:relax}.
The original derivation~\cite{AdGi} of
the AG equation relies on the assumption of the existence of correlated
regions of size $\xi$ rearranging cooperatively, 
and on the hypothesis that $S \xi^d \sim 
\textrm{const}$. As shown in  
Fig.~\ref{fig:1d}c, this relation seems to 
hold in the intermediate temperature window explored in the simulations, 
whereas it breaks down  below 
$T^{\star}$.

One  might  wonder  whether  the  whole  entropy,  $S$, which  we  have
 calculated,  plays the  role of  the ``configurational  entropy'' for
 supercooled  liquids   in  the   AG  relation.   When   $T<T_K$,  the
 thermodynamics is dominated  by the flipping of the  few single spins
 for  which the local  field is  comparable to  $T$. For  $T>T_K$, the
 thermodynamics is dominated by the excitation of ``droplets'' of size
 $\xi$.  Hence it  is  tempting  to regard  the  very low  temperature
 entropy in  our spin glass  simulations as being the  contribution to
 the entropy from a single state, and that the configurational entropy
 is that  which arises when  many droplets are thermally  excited.  We
 have studied  also the ``configurational entropy''  which is obtained
 by subtracting from  $S$ the linear contribution which  fits the very
 low  temperature entropy. However,   the same  qualitative results
  were found
 with this definition of $S$.

\begin{figure}[ht]
\vspace{-0.1cm}
\begin{center}
\psfig{figure=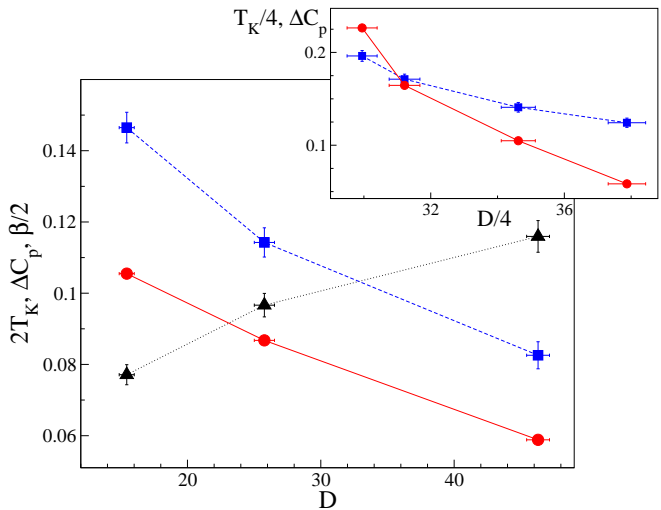,scale=0.31,angle=270}
\end{center}
\vspace{-0.4cm}
\caption{{\bf Main frame:} 
Kauzmann temperature $T_K$ (continuous line and circles), 
jump in the specific heat $\Delta C_p$ (dashed line and squares),
and stretching exponent $\beta$ (dotted line and triangles) as a function
of the fragility $D$ extrected from the VF fit of $\log \tau$ 
in $1d$. {\bf Inset:} Kauzmann temperature $T_K$ (continuous line and circles)
and jump in the specific heat $\Delta C_p$ (dashed line and squares)
as a function
of the fragility $D$ extracted from the VF fit of $\exp [4 \sigma_h^2 / T^2 ]$ \
in $3d$.}
\label{fig:frag}
\vspace{-0.3cm}
\end{figure}

Due to the success of the AG relation, 
it is natural to expect that the dynamics of
the system will be dominated by  the flipping of
spin domains of size  $\xi$. 
Since the variance of the the 
effective couplings decreases under iteration and approaches zero after 
a few iteration steps, whereas the effective magnetic fields 
are Gaussianly distributed with mean $h$ and variance $\sigma_h (T,h)$,
one might guess that the dynamics of the system is
equivalent to that of a chain of non-interacting
spin domains of linear extent proportional to $\xi$, sitting
in a random external field. Notice that this situation has been studied in 
Ref.~\cite{bouchaud}.
Taking into account the time to pass a spin flip through the domain, which
 involves the breaking of the largest bond in the domain whose
magnitude will be denoted by $L_i$,
the time to reverse  each spin domain
will be of the form $\tau \propto \exp [(2 L_i +2h_i)/T]$.
If the distribution of the $L_i$ is also Gaussian, 
the distribution of the sum $(L_i+h_i)$ will be another Gaussian of variance 
$\sigma_L^2 +\sigma_h^2T$.
According to Ref.~\cite{bouchaud}, 
this leads to the following expressions for the
relaxation time $\tau \simeq \exp[ 4 (\sigma_h^2+\sigma_L^2) / T^2]$
and for the stretching exponent $\beta \simeq C
[1 + 4(\sigma_h^2 + \sigma_L^2)/T^2]^{-1/2}$, with a constant $C=1$. 
We have verified that  
these formula work quite well in describing the dynamics of the system
in one dimension (with $\sigma_L \simeq 0.22$).
Nevertheless, they are not {\em perfect}: the constant
$C$ is bigger than one ($C \simeq 1.5$) and $\sigma_L$ is too small.
These discrepancies could be because the time it takes to flip the
 spins by breaking the
largest bond on a line of spins of length $\xi$ has not been handled with 
sufficient accuracy.
 The largest bond has
 its own probability distribution, which is just
not a Gaussian. A full theory would be complicated. However,
in $3d$  much bigger values 
of the external field have to be taken to keep the magnitude of the 
 correlation length  only a few lattice spacings at low temperatures,
 so one could reasonably expect that the relaxation time
 is dominated just
by the random fields alone.

\begin{figure}[ht]
\vspace{-0.1cm}
\begin{center}
\psfig{figure=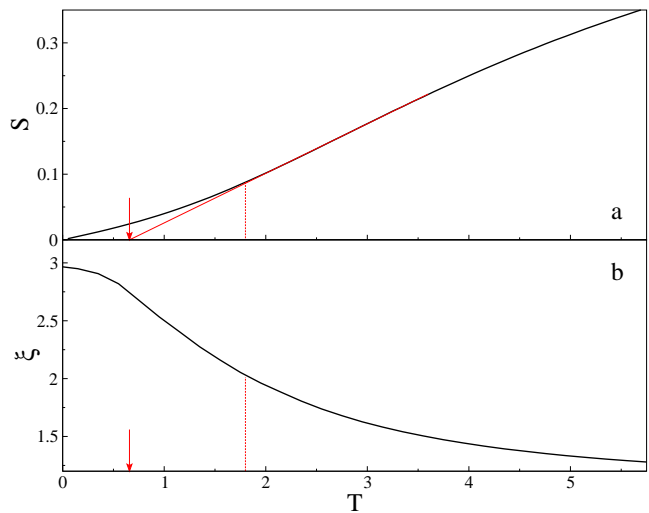,scale=0.31,angle=270}
\vspace{-0.4cm}
\end{center}
\caption{Three dimensional EA model within the MK approximation for $h=2.0$.
{\bf a)} Entropy per site, $S$; {\bf b)} Correlation length, $\xi$. The vertical 
arrows and the vertical dotted lines correspond, respectively, to $T_K$ 
and $T^{\star}$.}
\label{fig:3dentropy}
\vspace{-0.3cm}
\end{figure}

The fragility $D$,
and the Kauzmann temperature $T_K$ (see Eq.~(\ref{eq:vtf})) are both 
affected by the magnetic field, $h$. 
More precisely, $D$ decreases as $h$ is increased (i.e., stronger glasses 
are described by bigger fields), whereas $T_K$ (and $T^{\star}$) 
increases as $h$ is increased.
In the main frame
of Fig.~\ref{fig:frag}, the dependence of $T_K$ on the fragility is plotted.
We also plot the behavior of another two important quantities:
 the jump in the specific heat $\Delta C_p$ 
 and the stretching exponent $\beta$ (at $T=0.3$). 
The former can be estimated by (arbitrarily) setting $T_g \simeq T^{\star}$, 
so that $\Delta C_p \sim c(h) T^{\star}$, $c(h)$ being the slope of the 
entropy in the linear regime [see Eq.~(\ref{eq:1dentr})]; 
the latter can be directly computed from numerical simulations.

Interestingly enough, 
$\beta$ is found to be an increasing function of $D$, whereas
$\Delta C_p$ decreases as $D$ increases,
in agreement with  observations on supercooled liquids~\cite{LW}.

\section{Results in three dimensions} \label{sec:3d}

We now turn 
to the $3d$ case, which is  most 
relevant for real supercooled liquids.
 We have evaluated the  free energy of the   
 EA spin glass model in a field 
by means of the MK approximation, a real-space
renormalization group techniques that gives approximate recursion relations
for the flow of the coupling constants and magnetic fields 
distributions~\cite{kadanoff,MKA}.
 We have used 
the  ``bond moving'' procedure, where the bonds
on the $3d$ lattice are moved before each decimation step, so that no
higher order couplings are generated~\cite{MKA}:
in a $d$-dimensional lattice, $2^{d-1}$
bonds are superimposed and added up, whereas the ``naked'' spins that are left
behind have no couplings. Taking the trace over the spins that are on the
main bonds leads to the coupling constants, according to 
Eq.~(\ref{eq:1drg}), between neighboring spins on the coarse-grained lattice. 
The decimation procedure is iterated $n$ times on a lattice of size
$L = 2^n$.
There is a  flexibility in the MK renormalization scheme as to how
 the fields are 
moved. We have treated the field terms as belonging to bonds: when a bond
is moved we also move all its field terms to the end that is to be traced 
over next~\cite{MKA}.

In $3d$, within the MKA, 
when $h=0$ and $T<T_c$, $T_c\simeq 1.78$, the variance
 of the effective couplings 
grows indefinitely under iteration.  For finite
values of the magnetic field  there is no
evidence of a de Almeida-Thouless line~\cite{AlTh}. 
The variance of the couplings
might grow initially for low enough fields and/or temperatures, but  it always
decreases and eventually vanishes after a sufficient number of iteration 
steps, just like in $1d$.  
The average value of the field distribution equals the initial value
of the uniform magnetic field, $h$,  
whereas the width of the distribution saturates at a finite value, 
$\sigma_h (T,h)$.

The temperature dependence of 
the entropy per spin and the correlation length $\xi$,
obtained within the MKA from
 the exponential decay of
the variance of the effective coupling, $J_{ij}^{(n)}$,
which decreases as $\exp(-2^n/\xi)$ at large $n$,
are plotted in 
Figs.~\ref{fig:3dentropy}a and~\ref{fig:3dentropy}b for $h = 2.0$,
showing a scenario very similar to that found in $1d$:
there is a temperature range in which 
the entropy per site, $S$, is linear and is extrapolated to vanish 
at a finite 
Kauzmann temperature, $T_K$; 
 a crossover occurs at a higher temperature 
$T^{\star}$, where the entropy deviates from  linear behavior.
 The correlation length, $\xi$,
 grows as the 
temperature is decreased and approaches a finite value at $T=0$.

We also mention that, similarly to the $1d$ case, 
there is a modest range of values of the external magnetic field $h$ 
(for $1.8 \lesssim h \lesssim 2.5$) for which the AG relation, $S \xi^3
\sim \textrm{const}$,  holds in the temperature
region $T \gtrsim T^{\star}$. This AG relation,
however, breaks down below $T^{\star}$ and at high temperatures and in
contrast to the $1d$ case is of less utility. 

Due to the  magnitude  of the relaxation times, standard Monte Carlo
 simulations of 
the  $3d$ model are more challenging than in the $1d$ case, and
we will leave them for future investigations.
 However, since the values of interest of the magnetic field are much
bigger than in $1d$, one can surmise that
the flipping of a spin domain of size $\xi$ sitting in a random external 
magnetic field Gaussianly distributed and with variance 
$\sigma_h^2$ are the dominant dynamical processes; hence,
the energy barriers involved in such processes 
might account reasonably well for the system's relaxation time, 
leading to $\log \tau \simeq 4\sigma_h^2 / T^2$~\cite{bouchaud}.
Following this hypothesis, we have verified that
$ \exp [4\sigma_h^2 / T^2 ] $ can be well fitted
by a VF law with $T_{VF} = T_K$ 
for $T>T^{\star}$. From the VF fit of this quantity it is also possible
 to extract the 
fragility, $D$, in the $3d$ case. In the inset of Fig.~\ref{fig:frag}, 
$T_K$ and $\Delta C_p$ are plotted as a function of $D$, showing they have
 very similar behavior to that found in $1d$.

 The domain size $\xi$ according to
 droplet scaling is given by equating  the
cost of flipping a droplet of size $\xi$, $\xi^{\theta}$, to the field energy
which might be gained, $h\xi^{d/2}$. As in $3d$ the exponent $\theta$ is small, 
($\approx 0.2$)~\cite{mike}, it follows
 that $\sigma_h^2 \sim h^2\xi^{d} \sim const.$, which
implies that the AG relation $\log \tau \simeq A/TS$ should  hold.

 The stetching exponent $\beta$ would be expected to be
$[1+4\sigma_h^2/T^2]^{-1/2}$, provided again that the time taken to pass the
 domain wall through the domain is not significant.

 One feature of the MKA study in 3d is that the configurational entropy seems 
to be smaller than the quoted values near the glass transition, perhaps by
 as much as a factor of $3$ \cite{LW}. In the derivation of the mapping to
 spin glasses \cite{MoYe06} one can see that the field $h$ will be a
 function of both the temperature and density, rather than simply being a 
temperature independent constant as we have assumed here throughout
 for simplicity. Allowing for this
 temperature dependence  could significantly change
 the entropy. For example suppose
$h^2=h_0^2+b^2T^2$, then the high-temperature limit of the entropy is
$S=\ln[2\cosh b]$, and by adjusting $b$, can be made as large as desired.\\

\section{conclusions} \label{sec:concl}

The mapping between supercooled
liquids and spin glasses in an external magnetic 
field, proposed in Ref.~\cite{MoYe06}, thus seems to provide a
 semi-quantitative explanation
of the properties of supercooled liquids including 
 the Kauzmann paradox, the Vogel-Fulcher 
behavior of the relaxation time, the stretched exponential decays of 
correlation functions, 
 a growing lengthscale, and  the Adam-Gibbs relation in the  regime 
$T \gtrsim T_g$, which is the precursor regime accessed by the experiments,
where the correlation length is  growing with temperature
 but is still only a  few
intermolecular distances. The droplet scaling limit studied in 
Ref.~\cite{MoYe06}  is appropriate only when the  correlation length is
 much bigger than this.
   The
 large timescales which 
exist below the glass transition temperature $T_g$ prohibit the taking of
 equilibrium data below it and so the  apparent thermodynamic
and dynamical
singularities at $T_K$ cannot be accessed. In  our scenario, $T_K$ is
 only a crossover temperature at which
 the growing correlation length  saturates 
to a constant value. 

Financial support  by the European Community's Human Potential Programme
under contract HPRN-CT-2002-00307, DYGLAGEMEM, is acknowledged.

\end{document}